\documentstyle[preprint,aps]{revtex}
\widetext
\draft
\tighten

\begin{document}

\title{Addendum to ``COULOMB INTERACTION DOES NOT SPREAD INSTANTANEOUSLY"}

\bigskip

\author{{\bf  Rumen I. Tsontchev, Andrew E. Chubykalo and Juan M.
Rivera-Ju\'arez}}

\address {Escuela de F\'{\i}sica, Universidad Aut\'onoma de Zacatecas \\
Apartado Postal C-580\, Zacatecas 98068, ZAC., M\'exico\\
e-mails: {\rm rumen@ahobon.reduaz.mx} and {\rm andrew@ahobon.reduaz.mx}}


\thispagestyle{\empty}

\maketitle


\baselineskip 7mm

\begin{center}

Received $\;\;\;\;\;\;\;\;\;\;\;\;\;\;\;\;\;$ 2001

\end{center}

\begin{abstract}
We describe the results of experimental research by configurations changed
with respect to the original experimental configuration [1]. In this way
it is demonstrated that the basic part of the signal registered is due to
the Coulomb electric field and does not originate from the transversal
electromagnetic wave.

\end{abstract}
$$ $$ $$ $$ {\bf PACS}: 03.50.-z, 03.50.De

\clearpage

The basic doubt regarding the reliability of our results is due to the
fact that it is not known which part of the signal obtained by us is
due to the Coulomb interaction, and which part is due to the presence of
a regular transversal electromagnetic wave. The theoretical consideration
is difficult because of the complexity of the experimental construction.

For this reason, it is more appropriate to determine this by means of
experiment.

In our Coulomb electric field generator (Fig. 1) there are three
elements which emit transversal electromagnetic waves:

$$$$
$$$$
$$$$
$$$$
$$$$
$$$$
$$$$
$$$$
\begin{center}
FIG. 1. {\small Regular experimental configuration.}
\end{center}

\begin{quotation}

1. the currents which flow along the surface of the spheres of the Van
de Graaf generators

2. the spark, which jumps between the electrodes of the discharger

3. the currents which flow along the discharge cable
\end{quotation}

During our experiments, the discharge cable was electrically and
magnetically shielded. For its construction, the discharge cable was
heavy coaxial cable. The discharge current flowed along the central
conductor. As an electric shield, its shield with a thickness of
approximately 1 mm was used, which is much greater than the depth of
the skin layer (0.03 mm) for the resonant frequency of 14 Mhz. For the
magnetic shield, a 5 mm thick iron sectioned tube was used. The coaxial
cable was placed inside the iron tube without electric connection between
them. The discharger was not shielded.  The shielding of the spheres of
the Van de Graaf generators is not recommended as the signal due to the
Coulomb electric field would be lost.

Consequently, it is necessary to evaluate the contribution of the
transversal electromagnetic waves emitted both by the discharger and by
the Van de Graaf generator's spheres. In practice, it is only necessary to
consider the influence of the active Van de Graaf generator as the
passive generator is considerably further away from the antenna 2. For
this reason, the signal originating from the passive Van de Graaf
generator reaches antenna 2 (the measuring antenna) with a delay greater
than 60 ns. That is, the said signal reaches antenna 2 after measuring and
does not influence the experimental result.

The general idea of the proposed experiments is simple. If the initial
direction of the currents that flow through the discharger and along
the sphere's surface changes, then the sign of the first front of the
electromagnetic wave emitted by them must change. Consequently, if the
predominant part of the signal is due to the electromagnetic wave, the
sign of the first flank of the signal obtained by the antennas will
change. The amplitude will change simultaneously.

With this purpose, an orifice will be perforated in the head of the
active Van de Graaf generator. A thick ceramic wall tube, which does not
allow electric rupture up to a voltage of 150 kV (Fig.2), is placed in
the orifice.

$$$$
$$$$
$$$$
$$$$
$$$$
$$$$
$$$$
$$$$
\begin{center}
FIG. 2. {\small Modified head of the Van de Graaf active generator.}
\end{center}

A conductor, which is connected electrically to the other pole
of the sphere, is placed along the axis of the tube. As we will
demonstrate further on, a construction of this type allows for a
reversal of the currents along the surface of the sphere. It is important
to point out that the electromagnetic wave emitted from the current, which
passes through the central conductor, is isolated from the metallic walls
of the sphere.

Experiments have been carried out with four configurations. In fig.1 we
show a regular configuration; in figures 3, 4 and 5, the configurations
are modified. The direction of the initial movement of the electrons is
indicated by arrows in all of the figures.

$$$$
$$$$
$$$$
$$$$
$$$$
$$$$
$$$$
\begin{center}
FIG. 3. {\small Inverse direction configuration of the electric current
which passes along the surface of the head of the Van de Graaf active
generator.}
\end{center}

$$$$
$$$$
$$$$
$$$$
$$$$
$$$$
$$$$
$$$$
\begin{center}
FIG. 4. {\small Inverse direction configuration of the electric current
which passes through the discharger.}
\end{center}

$$$$
$$$$
$$$$
$$$$
$$$$
$$$$
$$$$
$$$$
\begin{center}
FIG. 5. {\small Inverse direction configuration of the electric current
which passe through the discharger and along the surface of the head of
the Van de Graaf active generator.}
\end{center}

Each configuration has been investigated in three regimes:

\begin{quotation}

a. the discharge cable is not shielded.

b. the discharge cable is a coaxial cable. The discharge current passes
through the central thread. The cable shield is not connected to earth.

c. as in b., but the shield is connected to earth.
\end{quotation}

In this way we can evaluate the influence of the discharge cable screen
on the processes. As we can see from the figures, the general flow of the
electrons is always in the same direction - from the passive Van de Graaf
generator towards the active Van de Graaf generator. Because of this,
the signal due to the Coulomb electric field in the case of the four
configurations does not undergo significant changes. Consequently, if the
basic part of the signal is due to the Coulomb interaction, we should
observe, in the case of the four configurations, an impulse with negative
first flank (Fig.6).

$$$$
$$$$
$$$$
$$$$
$$$$
$$$$
$$$$
$$$$
\begin{center}
FIG. 6. {\small Real form of the signal obtained by all of the
configurations.}
\end{center}

At the same time, in two configurations the direction of the current in
the discharger changes in the opposite way (Fig.4, Fig.5). In an
analogous manner, the direction of the currents which flow along the
surface of the sphere changes in two configurations (Fig.3, Fig.5).
Consequently, if our signal is due to the electromagnetic waves emitted,
or to the discharger, or the sphere, or both together, at least in one
configuration we should observe a change in the sign of the first flank on
the oscilloscope screen (Fig. 7).

$$$$
$$$$
$$$$
$$$$
$$$$
$$$$
$$$$
$$$$
\begin{center}
FIG. 7. {\small Supposed form of the signal generated from the
transversal electromagnetic wave by changed configuration.}
\end{center}

As the experiments are carried out we can see that ALWAYS, regardless of
the configuration and  measurement regime, an impulse as pointed out in
Fig.6 is obtained.

The proportion between the signal amplitudes, obtained in the
configurations presented in figures 3, 4 and 5, taking the signal
amplitude of the regular configuration as reference point (Fig. 1), has
been investigated. The result is that the relative deviation of the
amplitudes does not exceed 5\%.

Consequently, we can arrive at the definite conclusion that the dominant
part of our signal is due to the Coulomb interaction.  Consequently, if
the Coulomb interaction is propagated instantaneously, we should record
this fact. However, in reality we only record the propagation of a wave. In
this manner, we can consider that it has been demonstrated experimentally
that the Coulomb electric field is propagated like a Coulomb wave (i.e.
not instantaneously).

\end{document}